\documentclass[aps,prb,twocolumn,superscriptaddress,showpacs,preprintnumbers,amsmath,amssymb]{revtex4}

\usepackage{graphicx}
\usepackage{bm}

\begin{document}
\title{Inhomogeneous magnetism in La-doped CaMnO$_{3}$. (I)
Nanometric-scale spin clusters and long-range spin canting}

\author{E. Granado}
\affiliation{NIST Center for Neutron Research, National Institute
of Standards and Technology, Gaithersburg, Maryland 20899}
\affiliation{Center for Superconductivity Research, University of
Maryland, College Park, Maryland 20742}
\affiliation{Laborat\'{o}rio Nacional de Luz S\'{i}ncrotron, Caixa
Postal 6192, CEP 13084-971, Campinas, SP, Brazil}
\author{C. D. Ling}
\affiliation{Institut Laue-Langevin, BP 156, 38042 Grenoble Cedex
9, France}
\affiliation{Materials Science Division, Argonne
National Laboratory, Argonne, Illinois 60439}
\author{J. J. Neumeier}
\affiliation{Department of Physics, Montana State University,
Bozeman, MT 59717}
\author{J. W. Lynn}
\affiliation{NIST Center for Neutron Research, National Institute
of Standards and Technology, Gaithersburg, Maryland 20899}
\affiliation{Center for Superconductivity Research, University of
Maryland, College Park, Maryland 20742}
\author{D. N. Argyriou}
\affiliation{Materials Science Division, Argonne National
Laboratory, Argonne, Illinois 60439}

\begin{abstract}
Neutron measurements on Ca$_{1-x}$La$_{x}$MnO$_{3}$ ($0.00 \leq x
\leq 0.20$) reveal the development of a liquid-like spatial
distribution of magnetic droplets of average size $\sim 10$ \AA,
the concentration of which is proportional to $x$ (one cluster per
$\sim 60$ doped electrons). In addition, a long-range ordered
ferromagnetic component is observed for $0.05 \lesssim x \lesssim
0.14$. This component is perpendicularly coupled to the simple
$G$-type antiferromagnetic ($G$-AFM) structure of the undoped
compound, which is a signature of a $G$-AFM$+$FM spin-canted
state. The possible relationship between cluster formation and the
stabilization of a long-range spin-canting for intermediate doping
is discussed.

\end{abstract}


\maketitle

\section{INTRODUCTION}

Doped manganites are strongly correlated electron systems with
unusually large responses to external perturbations such as
magnetic field and pressure. While the most dramatic effects such
as colossal magnetoresistance have been observed in heavily-doped
compounds, systematic studies on lightly and moderately-doped
samples may reveal some fundamental aspects of manganite physics.
In these regimes, the antiferromagnetic (AFM) spin structures
shown by the undoped compounds tend to be destabilized by the
ferromagnetic (FM) exchange interactions mediated through the
charge carriers. Electron-doped CaMnO$_{3}$ samples are
particularly attractive model systems due to the relative
simplicity and chemical stability of the parent compound, which
shows a simple quasi-cubic crystal structure and an isotropic
$G$-AFM spin ground state \cite{Wollan}. For electron-doped
CaMnO$_{3}$, a relatively weak ferromagnetism has been observed up
to $\sim 15$ \% doping
\cite{Chiba,Maignan,Neumeier1,Neumeier2,Aliaga}. While the classic
de Gennes theory for lightly-doped manganites describes the weak
ferromagnetism in terms of spin-canted ground states
\cite{deGennes}, a number of more recent theoretical studies
indicate that homogeneous canted magnetic structures may not be
energetically stable, suggesting a tendency towards magnetic and
electronic phase segregation for both hole-doped
\cite{Zou,Yunoki1,Arovas,Yamanaka,Nagaev,Kagan,Yunoki2,Dagotto,Shen,Yi}
and electron-doped \cite{Yunoki2,Dagotto,Shen,Yi,Dunaevsky}
manganites. In fact, for moderately hole-doped LaMnO$_{3}$ ($5-8$
\% Ca- or Sr- doping), single crystal neutron-scattering studies
revealed the existence of nanometric-scale magnetic
inhomogeneities at low $T$ \cite{Hennion1,Hennion2,Hennion3}.
Whether electron-doped manganites actually mirror this effect is
an open experimental problem and a fundamental issue, since the
phase diagram of electron-doped manganites is in general
asymmetrical with respect to their hole-doped counterparts. For
instance, the ferromagnetic metallic ground state is not realized
for La-doped CaMnO$_{3}$, in stark contrast with the wide
compositional interval where this state is observed in Ca-doped
LaMnO$_{3}$.

Previous dc-magnetization \cite{Neumeier2,Cohn}, thermal
conductivity \cite{Cohn}, Raman-scattering \cite{Granado}, and
electron spin resonance \cite{Granado} studies on
Ca$_{1-x}$La$_{x}$MnO$_{3}$ indicate a crossover between distinct
doping regimes at $x \sim 0.03$, which in this paper we refer to
as low-doping ($0 < x \lesssim 0.03$) and intermediate-doping
($0.03 \lesssim x \lesssim 0.15$) regimes. While it has been
suggested that this crossover may reflect novel polaron physics
\cite{Cohn}, not much direct information on the microscopic
structure of the weak ferromagnetism observed for electron-doped
manganites is presently available. A notable exception is an NMR
study performed on Ca$_{1-x}$Pr$_{x}$MnO$_{3}$ ($x \leq 0.1$)
\cite{Savosta}, which found a coexistence of ferromagnetism and
antiferromagnetism in the samples studied, thus supporting a phase
segregation scenario.

In this work, the microscopic structure of the magnetic ground
states of ceramic pellets of Ca$_{1-x}$La$_{x}$MnO$_{3}$ ($x$ =
0.00, 0.02, 0.03, 0.05, 0.06, 0.07, 0.09, 0.12, 0.16, and 0.20)
are investigated by neutron measurements. We focus on the
compounds with $x=0.02$ and $x=0.09$, which are representative
members of the low- and intermediate-doping regimes, respectively.
Elastic scattering at low angles reveals a liquid-like spatial
distribution of magnetic clusters of average size $\sim 10$ \AA \
in both regimes, whose concentration in the $G$-AFM matrix is
proportional to the doping level. Diffraction measurements under
applied magnetic fields reveal that the $G$-AFM and FM spin
components are uncoupled for low-doping and become orthogonally
coupled as the doping increases. Such orthogonal coupling is a
signature of a spin-canted state. Small-angle neutron scattering
(SANS) measurements also show magnetic domain-wall scattering in
the intermediate doping regime, revealing a long-range FM
component. The combined results severely limit the possible
scenarios for the development of the FM moment in electron-doped
manganites. In fact, they indicate a non-trivial magnetism for
this system, which cannot be described either by a homogeneously
spin-canted state \cite{deGennes}, or by a radical phase
segregation where FM clusters are embedded into a pure $G$-AFM
matrix.

\begin{figure}
\includegraphics[width=0.5\textwidth]{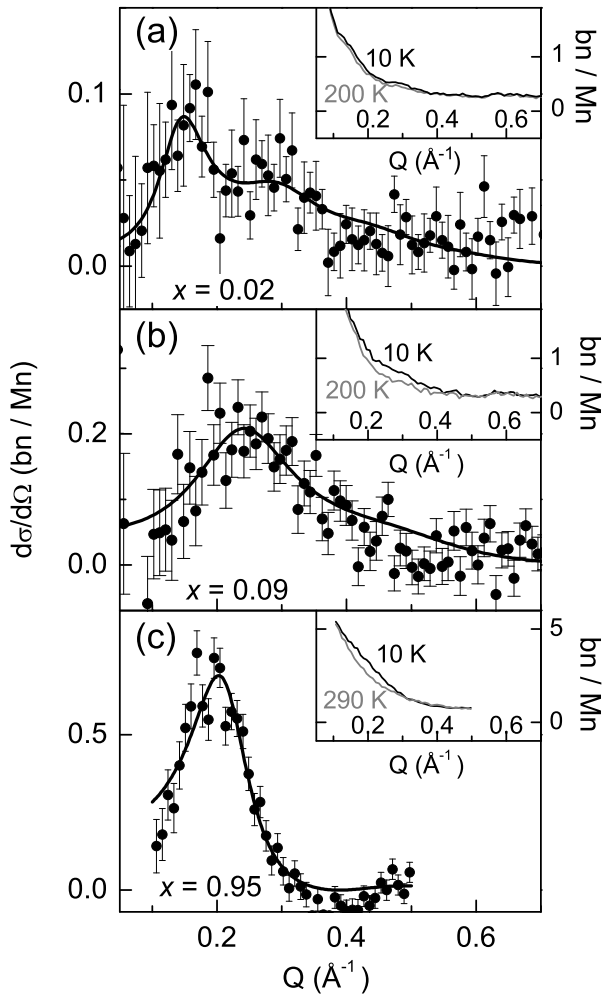}
\caption{\label{Clusters} Elastic magnetic cross section versus
$Q$ for Ca$_{1-x}$La$_{x}$MnO$_{3}$ for $x=0.02$ (a), $x=0.09$
(b), and $x=0.95$ (c). The solid lines are fits to a liquid-like
distribution model of magnetic droplets (see text). The insets
show the raw data at 10 K and 200 K (290 K for $x=0.95$).}
\end{figure}

\section{RESULTS AND ANALYSIS}

The La-doped CaMnO$_{3}$ samples were prepared by standard solid
state reaction, as described in detail in the following
paper.\cite{Ling} A hole-doped sample,
Ca$_{0.05}$La$_{0.95}$MnO$_{3}$, was prepared in a similar manner
to the other samples, but was reacted in Argon at all stages of
the preparation and reacted to a maximum temperature of 1250
$^{\circ}$C to keep the defect concentration low. Elastic neutron
scattering experiments at low angles were performed using the BT-2
triple-axis spectrometer at the NIST Center for Neutron Research,
with $E=14.7$ meV and $(60^{\prime }-20^{\prime }-20^{\prime
}-80^{\prime })$ collimation. The insets of Figs.
\ref{Clusters}(a) and \ref{Clusters}(b) show the elastic
scattering at 10 K and 200 K for Ca$_{1-x}$La$_{x}$MnO$_{3}$ with
$x=0.02$ and $0.09$, in the $Q$-interval between 0.05 \AA $^{-1}$
and 0.7 \AA $^{-1}$. The elastic magnetic scattering at low-$T$
($I_{M} (Q)$) can be more readily identified by subtracting the
elastic scattering at 200 K from the intensities at 10 K. This is
shown in Figs. \ref{Clusters}(a) and \ref{Clusters}(b) (symbols).
The solid lines are fits to a liquid-like model for the spatial
distribution of similar rigid magnetic droplets
\cite{Hennion2,Hennion3,Ashcroft}. The full expression for $I_{M}
(Q)$ under this model is given in ref. \cite{Hennion2}. The shape
of $I_{M} (Q)$ is determined by the minimum distance between
clusters ($d_{min}$), droplet diameter ($D$), and cluster
concentration ($N_{V}$). For $x=0.02$, the fitting parameters are
$d_{min}=41(3)$ \AA, $N_{V}=6.6(1.4)\cdot 10^{-6}$ \AA $^{-3}$
(i.e., one droplet per $\sim $59(12) doping electrons), and
$D=10.4(1.8)$ \AA. For $x=0.09$, we obtain $d_{min}=24(2)$ \AA ,
$N_{V}=28(6)\cdot 10^{-6}$ \AA $^{-3}$ (1 cluster per $63(14)$
doping electrons), and $D=10.6(1.6)$ \AA \ (see footnote
\cite{Comment}). Errors given in parentheses are statistical only
and represent one standard deviation. We note that fits to $I_{M}
(Q)$ of Fig. 1 assuming clusters with soft walls were also
performed, providing equally good fits to the experimental data
and nearly identical results for $N_{V}$ and $d_{min}$. In fact,
the calculated profiles shown in Figs. 1(a) and 1(b) are mostly
determined by inter-cluster diffraction, except for the overall
intensity decay at $Q \gtrsim 0.4$ \AA$^{-1}$ due to the finite
cluster size. Thus, little information on the cluster shape and
rigidity can be directly obtained from this experiment. Elastic
scattering experiments were also performed at 10 K and 290 K for a
polycrystalline hole-doped manganite,
Ca$_{0.05}$La$_{0.95}$MnO$_{3}$ (see Fig. 1(c)). The subtracted
intensity, $I(10$ K$)-I(290$ K), shows a peak at $Q \sim 0.2$
\AA$^{-1}$, for which the intensity, shape and width are in good
agreement with previously published results for a single crystal
of the same compound \cite{Hennion3}. This indicates that the
magnetic clusters observed for lightly hole-doped manganites
\cite{Hennion2,Hennion3} are essentially insensitive to the sample
growth method. This result, combined with the evidence for
magnetic clusters reported here for electron-doped manganites,
supports a universal tendency for intrinsic inhomogeneous ground
states in lightly-doped manganites.

\begin{figure}
\includegraphics[width=0.5\textwidth]{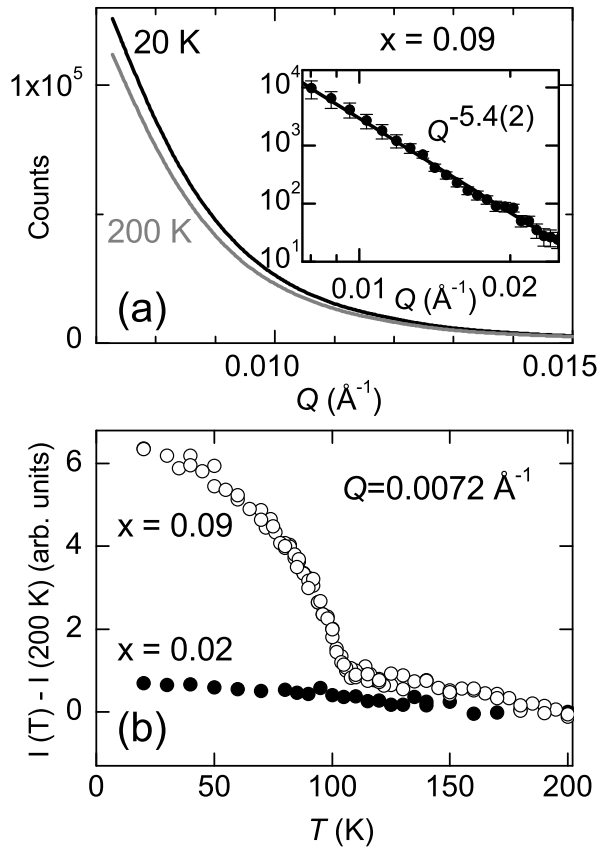}
\caption{\label{SANS}(a) SANS for Ca$_{1-x}$La$_{x}$MnO$_{3}$ at
20 K and 200 K for $x=0.09$. The inset shows I(20 K) - I(200 K)
and a fit to a power law. (b) $T$-dependence of the scattering at
$Q=0.0072$ \AA $^{-1}$ for $x=0.02$ and $x=0.09$. Data in (b) were
corrected for thickness and absorption to allow a direct
comparison between samples.}
\end{figure}

The possibility of a long-range FM component was investigated by
energy-integrated SANS. The FM scattering peaks at $Q=0$, and
shows a distribution in the $Q$-scale of $\sim 2 \pi /L_{d}$,
where $L_{d}$ is the average domain size. The experiments were
performed using the NG-1 instrument at NIST, with $\lambda =12$
\AA , and a sample-detector distance of 3.5 m. The intensities
were measured by a two-dimensional position-sensitive detector
(0.007 \AA $^{-1}<Q<$ 0.08 \AA $^{-1} $), and were angularly
averaged around the beam-center position.  Figure \ref{SANS}(a)
shows $I(Q)$ at $10$ K and $200$ K for $x=0.09$. Although the SANS
data are dominated by a non-magnetic and slightly $T$-dependent
component (most likely from intergrain scattering), a magnetic
component is also clearly present for $x\ge 0.05$. This is
evidenced by the $T$-dependence of the scattering at $Q=0.0072$
\AA $^{-1}$, showing a significant enhancement below $T_{C}$
($=108(1)$ K for $x=0.09$, see Fig. \ref{SANS}(b)). The inset of
Fig. \ref{SANS}(a) shows the intensities at 10 K after subtracting
the background scattering at 200 K, and a fit to a power-law
behavior, $I=AQ^{-5.4(2)}$, for $Q$ between 0.007 \AA$^{-1}$ and
0.025 \AA$^{-1}$. This result indicates the existence of magnetic
domains with sizes of several hundred Angstroms or larger,
evidencing a long-range FM component. This conclusion is also
supported by polarization-dependent neutron diffraction of a
nuclear Bragg peak for $x=0.09$, which showed the neutron beam
being depolarized by the sample below $T_{C}$ (not shown). For
$x=0.02$, no evidence for domain-wall scattering was observed by
SANS, within our experimental sensitivity (see Fig.
\ref{SANS}(b)).

Powder diffraction experiments were carried out over an extended
$Q$-range \cite{Ling}. For $x = 0.02$ and $0.03$, $G$-AFM Bragg
peaks were observed at low $T$; the weak FM component seen by
dc-magnetization \cite{Neumeier2} was below our experimental
sensitivity. For $0.06 \leq x \leq 0.12$, $G$-AFM Bragg peaks and
FM intensities on top of nuclear Bragg peaks were observed at
low-$T$, as well as magnetic reflections from the $C$-type AFM
structure ($C$-AFM) \cite{Wollan}. The FM Bragg intensities
confirm the existence of a spontaneous long-range FM component in
the intermediate-doping regime, in accordance with our SANS
measurements (see above). For $x=0.16$ and $0.20$, only $C$-AFM
magnetic Bragg peaks were observed. A combined analysis using
high-resolution neutron and synchrotron X-ray diffraction data
makes it clear that the $C$-AFM magnetic reflections originate in
crystallographic domains having a distinct (monoclinic) crystal
structure due to the elongation of MnO$_{6}$ octahedra along the
$C$-AFM chain direction. Such regions coexist on a mesoscopic
scale with orthorhombic domains possessing regular MnO$_{6}$
octahedra. Details are given in the following paper.\cite{Ling}

To clarify the microscopic relationship between the FM signal and
the $G$-AFM and $C$-AFM spin components, $H$-dependent neutron
diffraction experiments were carried out on BT-2 with
$E_{i}=E_{f}=14.7$ meV and $(60^{\prime }-40^{\prime }-40^{\prime
}-open)$ collimation. The field was applied perpendicularly to the
plane defined by the incident and scattered wave vectors, using a
superconducting magnet. The magnetic intensities are proportional
to the square of the sublattice magnetization, and
also to the geometrical factor $\gamma \equiv \left\langle 1-(%
\widehat{{\bf M}}\cdot \widehat{{\bf \tau }})^{2}\right\rangle $,
where $\widehat{{\bf \tau }}$ and $\widehat{{\bf M}}$ are the
directions of the reciprocal lattice vector and the sublattice
magnetization, respectively, and the brackets account for a
domain-average. For cubic or quasi-cubic crystal lattices, $\gamma
({\bf H}=0)=2/3$. Under the application of ${\bf H}$, the FM
component reorients along the field direction. Therefore, for
increasing ${\bf H}\perp \widehat{{\bf \tau }}$, such as in our
experiment, one has $\gamma _{FM}({\bf H})\rightarrow 1$. The
coupling of the AFM to the FM moments can be inferred from the
${\bf H}$-dependence of $\gamma _{AFM}$, as described below (see
also refs. \cite{Wollan} and \cite{Granado2}).

\begin{figure}
\includegraphics[width=0.50\textwidth]{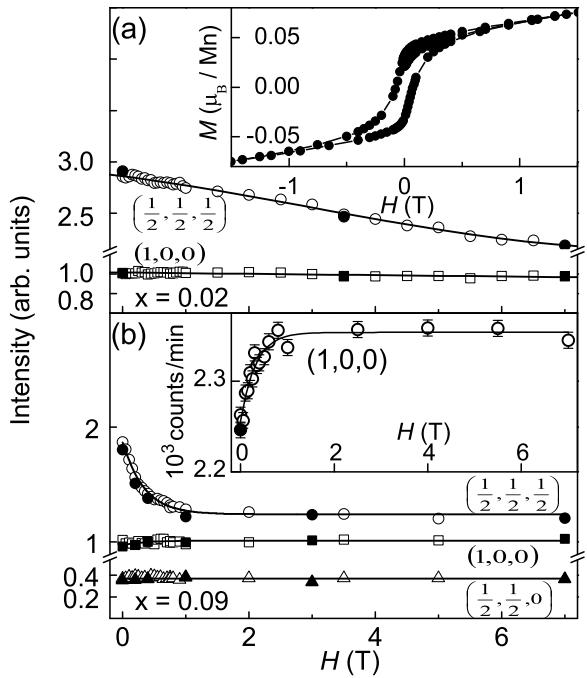}
\caption{\label{Mag}$H$-dependence at 5 K of the intensity of the
$(\frac{1}{2},\frac{1}{2},\frac{1}{2})$ $G$-AFM and $(1,0,0)$
nuclear + FM Bragg reflections for $x=0.02$ (a) and $x=0.09$ (b),
and $(\frac{1}{2},\frac{1}{2},0)$ $C$-AFM reflection for $x=0.09$.
Empty (filled) symbols represent increasing (decreasing) fields.
The insets show the $H$-dependence at 5 K of the dc-magnetization
for $x=0.02$ and of the peak intensity of the $(1,0,0)$ Bragg
reflection for $x=0.09$.}
\end{figure}

Figure \ref{Mag}(b) shows the field-dependence of the $(1,0,0)$
nuclear + FM, $(\frac{1}{2},\frac{1}{2},\frac{1}{2})$ $G$-AFM, and
$(\frac{1}{2},\frac{1}{2},0)$ $C$-AFM Bragg peaks (cubic notation)
for $x=0.09$. The inset of Fig. 3(b) shows in detail the peak
intensity of the $(1,0,0)$ reflection. The observed increase of
this peak intensity for increasing fields up to $\sim 0.5$ T
indicates a reorientation of the FM spin-component along the field
direction (see above). The intensity of the
$(\frac{1}{2},\frac{1}{2},\frac{1}{2})$ peak decreases by 34(3) \%
in the same field range, indicating a perpendicular coupling
between $G$-AFM and FM spin components, consistent with $G$-AFM +
FM spin-canting. The intensity of the
$(\frac{1}{2},\frac{1}{2},0)$ peak is insensitive to fields up to
7 T, showing that the $C$-AFM spin component is not coupled to the
FM spin component. Thus, the results shown in Fig. 3(b), combined
with high-resolution diffraction data \cite{Ling}, suggest
mesoscopic phase coexistence between C-AFM regions with no FM
moment and $G$-AFM + FM state for $0.06 \lesssim x \lesssim 0.12$.

For $x=0.02$, the field-induced reorientation of the very weak FM
spin component was probed by dc-magnetization ($M_{dc}$)
measurements, which were taken using a commercial SQUID
magnetometer. The inset of Fig. \ref{Mag}(a) shows the
$H$-dependence of $M_{dc}$ at 5 K. The curve can be decomposed
into a FM signal which saturates at $\sim 0.05 \mu_{B}$/Mn for
fields smaller than 0.5 T, and a linear component which is
tentatively ascribed to a conventional field-induced spin canting.
The field-dependence of the $G$-AFM spins for $x=0.02$ was probed
by neutron diffraction (see Fig. \ref{Mag}(a)). An intensity
decrease of the $(\frac{1}{2},\frac{1}{2},\frac{1}{2})$ reflection
was observed in the field scale of several tesla. This effect is
not directly connected to the reorientation of the spontaneous FM
moments, which takes place for $H<0.5$ T (see inset of Fig.
\ref{Mag}(a)). Thus, for $x=0.02$, the $G$-AFM moments are not
coupled to the FM moments, at least for small fields ($H<0.5$ T),
and the origin of the weak FM signal for this compound is
inconsistent with a zero-field spin-canting of the $G$-AFM
structure.

\section{DISCUSSION}

The observation of magnetic clusters (see Fig. 1) clearly points
to a spatially inhomogeneous charge-carrier distribution in this
system. The ratio between doped electrons and cluster densities
($\sim 60$, see above) is independent of $x$ for electron-doped
manganites and is identical to that found in hole-doped manganites
\cite{Hennion2,Hennion3}, strongly suggesting a universal
behavior. However, this large ratio and the small dimensions of
the observed clusters (comprising $\sim 10$ unit cells) make it
clear that only a fraction of the doped electrons are inside such
clusters. The correct mechanism that leads to this phenomenon is
not clear at this point. Even with a few electrons in each
cluster, the charge contrast inside and outside the droplets may
be exceedingly high, particularly in the low-doping regime. Simple
electrostatic considerations indicate that the Coulomb energy loss
for a FM two-electron droplet with $D \sim 10$ \AA\ surrounding a
La$^{3+}$ ion is of the order of 1 eV for low- and
intermediate-doping regimes, and increases quadratically with the
number of cluster electrons. This Coulomb energy might overwhelm
the delocalization energy gain per electron in the cluster ($t
\sim 0.1-1$ eV), as already pointed out by Chen and Allen
\cite{Chen1}. In this context, it would appear natural to consider
that clusters might be formed by electrostatic attraction in
La-rich regions of the sample, presumably associated with
intrinsic chemical inhomogeneities \cite{Shibata}. This mechanism
would lead to electrically-neutral, Mn$^{3+}$-rich, magnetic
clusters. The relatively small cluster densities would be
naturally accounted for in this scenario. On the other hand, the
cluster diffraction profiles shown in Figs. 1(a) and 1(b) imply a
spatial short-range order similar to a liquid state, as opposed to
a cluster gas where the cluster positions would be uncorrelated.
Such an order suggests intercluster repulsion, presumably dictated
by Coulomb forces between electrically charged and mobile
clusters. The cluster diffraction also implies that neighboring
clusters are magnetically correlated in both low- and
intermediate-doping regimes, as opposed to a superparamagnetic
state. In view of the above considerations, we believe that a
truly intrinsic mechanism for small cluster formation in this
system, i.e., not caused by chemical inhomogeneities, should not
be discarded at this point.

The electrons outside the small magnetic clusters discussed above
are likely to be important for the overall magnetic behavior of
La-doped CaMnO$_{3}$. In fact, using the fitting parameters
obtained from Fig. 1, the total cluster contributions to the
sample-average magnetizations are estimated to be $0.02(1)
\mu_{B}/$Mn for $x=0.02$ and $0.04(2) \mu_{B}/$Mn for $x=0.09$,
which are significantly smaller than the saturation magnetizations
obtained from dc-magnetometry, $0.05$ and $0.40 \mu_{B}/$Mn,
respectively \cite{Neumeier2}. Also, the combination of a
long-range FM spin component and the orthogonal coupling between
FM and $G$-AFM spin components at intermediate-doping is a
signature of a long-range $G$-AFM + FM spin-canted state that does
not appear to be accomplished at the low-doping regime. Although
the present set of experimental data, combined with previous work
on La-doped CaMnO$_{3}$ \cite{Neumeier1,Neumeier2,Granado,Cohn},
may be insufficient to lead to a complete description for the
microscopic structure of the FM moments and doped electrons in
this system, it severely constrains any plausible model, as
described below.

It is clear from the results above that a second type of doped
electron is present in La-doped CaMnO$_{3}$, besides the type
forming relatively small FM clusters ($D \sim 10$ \AA). Given the
long-range spin-canted state evidenced for intermediate doping,
the extra electrons seem to be delocalized on the atomic scale. On
the other hand, the fact that a metallic state is not accomplished
at low temperatures \cite{Neumeier1}, combined with the absence of
an observable long-range FM component at low-doping, suggests that
such extra electrons are not fully delocalized into a de Gennes
canted state \cite{deGennes} either. Thus, we suggest that these
electrons are segregated into spin-canted regions of finite size,
presumably larger than the small FM clusters directly observed by
neutrons. These regions would overlap for intermediate-doping,
leading to the observed long-range FM component perpendicularly
coupled to the $G$-AFM moments. We note that such hypothetical
spin-canted clusters were not directly observed in our neutron
scattering measurements, possibly due to the small magnetization
contrast and/or large sizes leading to small differential cross
section in the $Q$-region accessible for elastic measurements (see
Fig. 1). From a theoretical point of view, the formation of an
inhomogeneous $G$-AFM$+$FM spin-canted state in electron-doped
manganites, evidenced in this work, might be the result of a
balance between the well-known electronic instability of the
homogeneous spin-canted $G$-AFM state
\cite{Yunoki2,Dagotto,Shen,Yi,Dunaevsky} and the large Coulomb
energy cost of a radical phase segregation scenario where purely
FM droplets are formed into a pure $G$-AFM background.

\section{CONCLUSIONS}

Our results on La-doped CaMnO$_{3}$ indicate that a fraction of
the doped electrons segregate into small ($D \sim 10$ \AA) FM
clusters embedded in the $G$-AFM matrix of the undoped compound.
The remaining electrons are presumably delocalized over a more
extended volume, leading to an inhomogeneous spin-canted state at
intermediate doping. The density of the $10$ \AA -clusters, as
well as the FM component of the spin-canted state, increase with
the doping level, and the overall FM moment becomes increasingly
dominant over the $G$-AFM spin component. Nevertheless, the pure
FM metallic state is never stabilized for La-doped CaMnO$_{3}$,
due to the gradual emergence of the orbitally-ordered $C$-AFM
state for $x \gtrsim 0.06$, which competes with the $G$-AFM + FM
state through a first-order phase transition, as explored in the
following paper \cite{Ling}. This competition leads to mesoscopic
magnetic and crystallographic phase separation over a large $x$-
and $T$-interval, and finally to the stabilization of the $C$-AFM
phase for $0.16 \lesssim x \lesssim 0.20$
\cite{Ling,Santhosh,Pissas1,Pissas2}.

\section{Acknowledgements}

This work was supported by FAPESP, Brazil, NSF-MRSEC, DMR
0080008, NSF DMR 9982834, and US Department of Energy, Basic
Energy Sciences - Materials Sciences, contract W-31-109-ENG-38,
USA.


\begin{references}
\bibitem{Comment} An estimation of the difference between the
average magnetization inside and outside the droplets ($\delta M$)
may be obtained from the cross sections, as described in ref.
\cite{Hennion2}. We obtain $\delta M=6(3)$ $\mu _{B}/$Mn for
$x=0.02$ and $x=0.09$. The large standard deviation is mostly due
to the uncertainty in the cluster volume. For $x=0.09$, we
considered that the droplets are distributed in the $\sim 50$ \%
volume fraction occupied by the G-AFM + FM mesoscopic domains (see
text and Ref. \cite{Ling}). Thus, the actual differential cross
sections for droplet scattering from such mesoscopic domains is
twice the sample-average cross sections shown in Fig. 1(b). The
fitting parameter $N_{V}$ refers, by definition, to the cluster
concentration inside the mesoscopic domains where the clusters are
located.
\end{references}
\end{document}